\documentclass[12pt]{spieman}  
\usepackage{amsmath,amsfonts,amssymb}
\usepackage{graphicx}
\usepackage{setspace}
\usepackage{tocloft}

\usepackage{siunitx}
\usepackage{booktabs}
\usepackage{lineno}

\newcommand{\nm}{\,\si{nm}}
\newcommand{\um}{\,\si{{\micro}m}}
\newcommand{\mm}{\,\si{mm}}
\newcommand{\cm}{\,\si{cm}}
\newcommand{\m}{\,\si{m}}
\newcommand{\km}{\,\si{km}}
\newcommand{\eV}{\,\si{eV}}
\newcommand{\keV}{\,\si{keV}}

\title{Multi-Image X-ray Interferometer Module: I. design concept and proof-of-concept experiments with fine-pitch slits}

\author[a]{Kazunori Asakura}        
\author[a,b,c]{Kiyoshi Hayashida} 
\author[d]{Tomoki Kawabata}        
\author[c]{Yoneyama Tomokage}  
\author[a,b]{Hirofumi Noda}          
\author[a,b]{Hironori Matsumoto}  
\author[a]{Hiroshi Tsunemi}           
\author[e]{Hiroshi Nakajima}         
\author[f]{Hisamitsu Awaki}          
\author[g]{Junko S. Hiraga}            

\affil[a]{Department of Earth and Space Science, Graduate School of  Science, Osaka University, 1-1 Machikaneyama, Toyonaka, Osaka 560-0043, Japan}
\affil[b]{Project Research Center for Fundamental Sciences, Graduate School of Science, Osaka University, 1-1 Machikaneyama, Toyonaka, Osaka 560-0043, Japan}
\affil[c]{Institute of Space and Astronautical Science, Japan Aerospace Exploration Agency, 3-1-1 Yoshino-dai, Chuo-ku, Sagamihara, Kanagawa 252-5210, Japan}
\affil[d]{KIOXIA Corporation, 2-5-1 Kasama, Sakae-ku, Yokohama, Kanagawa 247-8585, Japan}
\affil[e]{College of Science and Engineering, Kanto Gakuin University, 1-50-1 Mutsuurahigashi, Kanazawa-ku, Yokohama, Kanagawa 236-8501, Japan}
\affil[f]{Department of Physics, Ehime University, 2-5 Bunkyo-cho, Matsuyama, Ehime 790-8577, Japan}
\affil[g]{Department of Physics, Kwansei Gakuin University, 2-1 Gakuen, Sanda, Hyogo 669-1337, Japan}

\cftpagenumbersoff{figure}
\cftpagenumbersoff{table} 
\begin{document} 
\maketitle

\begin{abstract} 

We propose a novel x-ray imaging system, Multi-Image X-ray Interferometer Module (MIXIM), with which a very high angular resolution can be achieved even with a small system size.
MIXIM is composed of equally-spaced multiple slits and an x-ray detector, and its angular resolution is inversely proportional to the distance between them.
Here we report our evaluation experiments of MIXIM with a newly adopted CMOS sensor with a high spatial resolution of $2.5\um$.
Our previous experiments with a prototype MIXIM were limited to one-dimensional imaging, and more importantly, 
the achieved angular resolution was only $\sim\ang{;;1}$, severely constrained due to the spatial resolution of the adopted sensor with a pixel size of $4.25\um$.
By contrast, one-dimensional images obtained in this experiment had a higher angular resolution of $\ang{;;0.5}$ when a configured system size was only $\sim1\m$,
which demonstrates that MIXIM can simultaneously realize a high angular resolution and compact size.
We also successfully obtained a two-dimensional profile of an x-ray beam for the first time for MIXIM by introducing a periodic pinhole mask.
The highest angular resolution achieved in our experiments is smaller than $\ang{;;0.1}$ with a mask-sensor distance of $866.5\cm$, which shows the high scalability of MIXIM. 
\end{abstract}


{\noindent \footnotesize\textbf{*}Kazunori Asakura,  \linkable{asakura\_k@ess.sci.osaka-u.ac.jp} }

\begin{spacing}{2}   

\section{Introduction}
\label{sec:intro}  

In the history of x-ray astronomy, the quality of x-ray imaging, in particular the angular resolution, has been a key factor.
Among the x-ray astronomical satellites launched so far, \textit{Chandra} X-ray Observatory (launched in 1999) 
has the highest angular resolution of $\ang{;;0.5}$, realized with a combination of Wolter type I mirrors and x-ray CCDs \cite{Weisskopf2000}.
The fact that \textit{Chandra} has produced a number of significant discoveries to this day (see e.g., Ref. \citenum{Tananbaum2014}) demonstrates the importance of a high angular resolution.
The angular resolution of the \textit{Chandra}-type imaging system is mainly determined by the degree of the surface smoothness and alignment accuracy of the mirrors,
which are achieved only through costly high-precision manufacturing.
Given the enormous cost in manufacturing the \textit{Chandra} mirrors (estimated to be several hundred million dollars\cite{ODell2012}),
it is unfeasible to further improve the angular resolution from that of \textit{Chandra}, or even to match it, with similar methods.

Meanwhile, other approaches have been proposed to achieve a high angular resolution, most notably x-ray interferometry.
\textit{MAXIM} and \textit{MAXIM Pathfinder}\cite{Cash2003} are pioneering x-ray interferometry projects with the goals of angular resolutions of $0.1$ micro-arcseconds and $100$ micro-arcseconds, respectively.
Their prototype interferometer has already succeeded in detecting interference fringes with an angular resolution of $\ang{;;0.1}$ for $1.25\keV$ x-rays in a ground experiment\cite{Cash2000}.
While it appears excellent, their major downsize is the technical difficulty for the actual deployment on a satellite.
Their system requires a very large size to make interference fringes resolvable with detectors (likely CCDs) in terms of the distance between mirrors and detectors,
exacerbated further by the fact that the incident grazing angle needs to be increasingly smaller for higher incident x-ray energies.
Even the above-mentioned ``small" prototype had a distance between mirrors and an x-ray CCD of $100\m$.
Indeed, the proposed design of \textit{MAXIM Pathfinder} has a mirror aperture size of $\sim1\m$ and specifies that it must maintain a distance of $200\km$ between the mirrors and detectors\cite{Gendreau2003}.
Although a formation flight of multiple satellites has been proposed to achieve a system with such a large size,
it is technically very challenging, and so far no projects to deploy \textit{MAXIM} or similar systems in orbit are on the horizon.

Several alternative designs of x-ray interferometry systems that considerably reduce the required size of the system, in particular the distance between the optics and detectors,
have been proposed (e.g., an interferometry system with a slatted mirror\cite{Willingale2004} or with a multi-layer beam splitter\cite{Kitamoto2011, Kitamoto2014}).
A system with a slatted mirror is also discussed in a future design concept of an x-ray interferometer in Voyage 2050 (the next planning cycle of the ESA Science Programme)\cite{Uttley2019}.
In theory, an interferometer with a slatted mirror and a typical x-ray CCD with a system size of $\sim20\m$ is expected to have an angular resolution of sub-milliarcseconds for $1.24\keV$ x-rays. 
In reality, however, no team has so far reported a successful detection of interference fringes in an x-ray band,
presumably due to the difficulty in manufacturing slatted mirrors within the very high required precision for its shape and surface smoothness\cite{Willingale2005, Willingale2013}.
As such, it has been a serious challenge to realize an x-ray imaging system with a higher angular resolution than that of \textit{Chandra} and with the scale of a single satellite ($< 10\m$).

Under these circumstances, we proposed Multi-Image X-ray Interferometer Module (MIXIM) in 2016\cite{Hayashida2016} as a system for high-resolution x-ray imaging, applying a novel imaging principle.
MIXIM does not require the precise mirrors as opposed to \textit{Chandra} or the aforementioned x-ray interferometers and has only the size of a micro-satellite (typically $50\cm$),
yet it has potential to realize a higher angular resolution than that of \textit{Chandra} for $12.4\keV$ x-rays.
Small-scale missions including micro-satellites have a strong advantage in feasibility (e.g., fabrication time, costs, and launch opportunity).
We have already demonstrated that a proto-type model of MIXIM achieved an angular resolution of about $\ang{;;1}$ for $12.4\keV$ albeit one-dimensional imaging\cite{Hayashida2018}.
Furthermore, the angular resolution could have been considerably improved with a detector of a higher spatial resolution because that was the main limiting factor for its imaging performance.
A type of new scientific CMOS sensor with a much higher spatial resolution was recently released commercially\cite{Yokoyama2018}.
Here we adopted it to MIXIM and evaluated the imaging performance of the renewed MIXIM system, expecting a significant improvement in the imaging capability from our previous experiment.

In this paper, we report the basic concepts of MIXIM in section \ref{sec:concepts} and our proof-of-concept experiments with the new detector and results in sections \ref{sec:experiment} and \ref{sec:results}, respectively.
Discussion of features of MIXIM in comparison with the other existing imaging systems and potential application of MIXIM to x-ray astronomy is given in section \ref{sec:discussion}.

\section{Concepts of MIXIM}
\label{sec:concepts}  

\subsection{Principle}
\label{sec:concepts_1}

\begin{figure} [ht]
   \begin{center}
   \begin{tabular}{c} 
   \includegraphics[height=7cm]{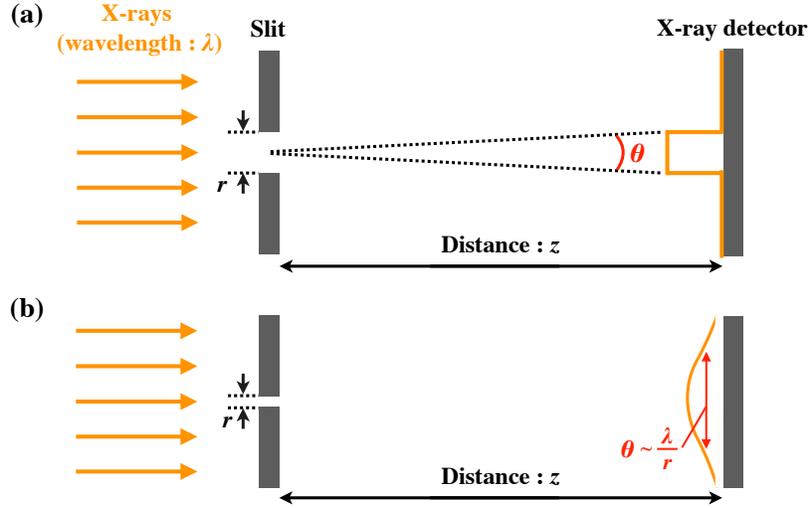}
   \end{tabular}
   \end{center}
   \caption[Configuration of the single slit camera.]  
   { \label{fig:SlitCamera}  
    (a) Configuration of the single slit camera with a slit width of $r$. (b) Same as (a), but in the case where diffraction dominates.}
   \end{figure}

The imaging principle of MIXIM is basically the same as that of a slit camera used in some x-ray astronomical imaging systems 
(e.g., MAXI/GSC\cite{Mihara2011, Sugizaki2011}, MAXI/SSC\cite{Tsunemi2010, Tomida2011}). 
Figure \ref{fig:SlitCamera} (a) depicts the slit camera with a slit width of $r$ and distance between the slit and detector of $z$.
With these notations, the angular resolution $\theta$ is approximated to be $rz^{-1}$ where $r \ll z$.
Although the equation predicts that making $r$ of the system smaller should unlimitedly improve the angular resolution,
in reality, diffraction comes into effect as a function of the wavelength of the incoming light, which is increasingly significant for smaller $r$ 
and limits the best-possible angular resolution to $\theta \sim \lambda r^{-1}$, as illustrated in Fig. \ref{fig:SlitCamera} (b).

\begin{figure} [ht]
   \begin{center}
   \begin{tabular}{c} 
   \includegraphics[height=7cm]{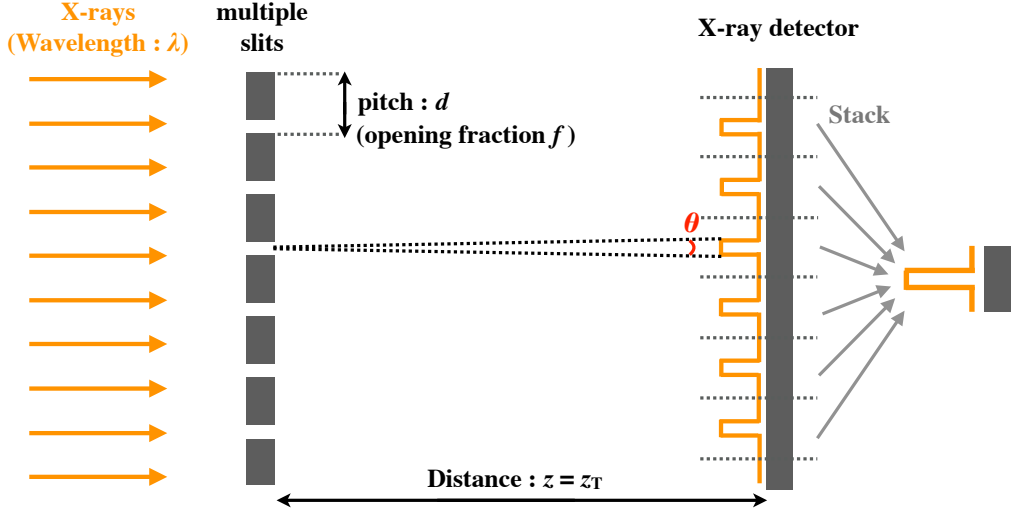}
   \end{tabular}
   \end{center}
   \caption[Configuration of the MIXIM.]  
   { \label{fig:MIXIM_principle}  
     Schematic drawing of the concept applied to MIXIM. The angular resolution of the system is determined primarily by the distance $z_{\mathrm{T}}$ and grating aperture size $fd$.
     The Talbot effect cancels the diffraction effect illustrated in Fig. \ref{fig:SlitCamera}.}
   \end{figure}

In MIXIM, we utilized the Talbot effect\cite{Talbot1836, Rayleigh1881}. 
Incident monochromatic parallel light passing through regular-interval apertures forms self-images at the distance $z_{\mathrm{T}}$ according to the following formula:\cite{Wen2013}
\begin{linenomath}
\begin{align}
  \label{eq:distance_plane}
  z_{\mathrm{T}} = m\frac{d^2}{\lambda} \,(m = 1, 2, 3...),
\end{align}
\end{linenomath}
where $d$ is the pitch of the apertures and $\lambda$ is the wavelength of the incident light (the self-images are shifted by $d/2$ for odd integers).
This interference phenomenon is called the Talbot effect and has been applied in various fields of optics, including x-ray imaging\cite{Momose2003, Pfeiffer2006} (summarized in e.g., Ref. \citenum{Wen2013}).
We applied it to MIXIM to circumvent the diffraction problem;
when we use equally-spaced multiple slits with a pitch of $d$ and opening fraction of $f$ instead of the single slit and adjust the distance $z$ to satisfy Eq. \ref{eq:distance_plane} 
for the wavelength $\lambda$ of our interest (hereafter referred to as the target wavelength), the light passing through each slit always forms the image of the x-ray source 
in the same way as in the ideal situation without diffraction with the slit camera even when the slit width $r$ is very small (as illustrated in Fig. \ref{fig:MIXIM_principle}).
Stacking all the images then yields a single reproduced image with high photon statistics which represents the x-ray source profile at the target wavelength (n.b., in reality, the image is convolved with the aperture pattern).

Whereas this imaging principle assumes monochromatic incoming x-rays, in reality, celestial objects usually have a broad band of x-rays.
Hence, MIXIM adopts a photon-counting x-ray detector with a spectroscopic capability for extracting only x-rays with the target wavelength.
With the aforementioned configuration of MIXIM (Fig. \ref{fig:MIXIM_principle}), the angular resolution is given by
\begin{linenomath}
\begin{align}
  \label{eq:resolution}
  \theta = \frac{fd}{z_{\mathrm{T}}} = \ang{;;0.4} \left( \frac{f}{0.2} \right)  \left( \frac{d}{5\um} \right) \biggl(\frac{z_{\mathrm{T}}}{50\cm} \biggl)^{-1}.
\end{align}
\end{linenomath}
With the pitch $d$ of $5\um$ and distance $z_{\mathrm{T}}$ of $50\cm$, x-rays with a wavelength of $0.1\nm$ ($12.4\keV$) satisfy Eq. \ref{eq:distance_plane};
in this case, we can rewrite Eq. \ref{eq:resolution} with the wavelength $\lambda$ and a positive integer $m$ as
\begin{linenomath}
\begin{align}
  \label{eq:resolution2}
  \theta = \frac{f\lambda}{dm} = \ang{;;0.4} \left( \frac{f}{0.2} \right) \left( \frac{\lambda}{0.1\nm} \right) \left( \frac{d}{5\um} \right)^{-1} \biggl( \frac{m}{2} \biggl)^{-1}.
\end{align}
\end{linenomath}
These equations indicate that the theoretical angular resolution of equally-spaced multiple slits with a pitch of few \si{{\micro}m} surpasses that of \textit{Chandra} at $12.4\keV$
and that it is achievable with a system that can be deployed on the typical size of micro-satellites.

\subsection{Calculation of Diffraction Patterns}
\label{sec:concepts_2}

\begin{figure} [ht]
   \begin{center}
   \begin{tabular}{c} 
  \includegraphics[height=8cm]{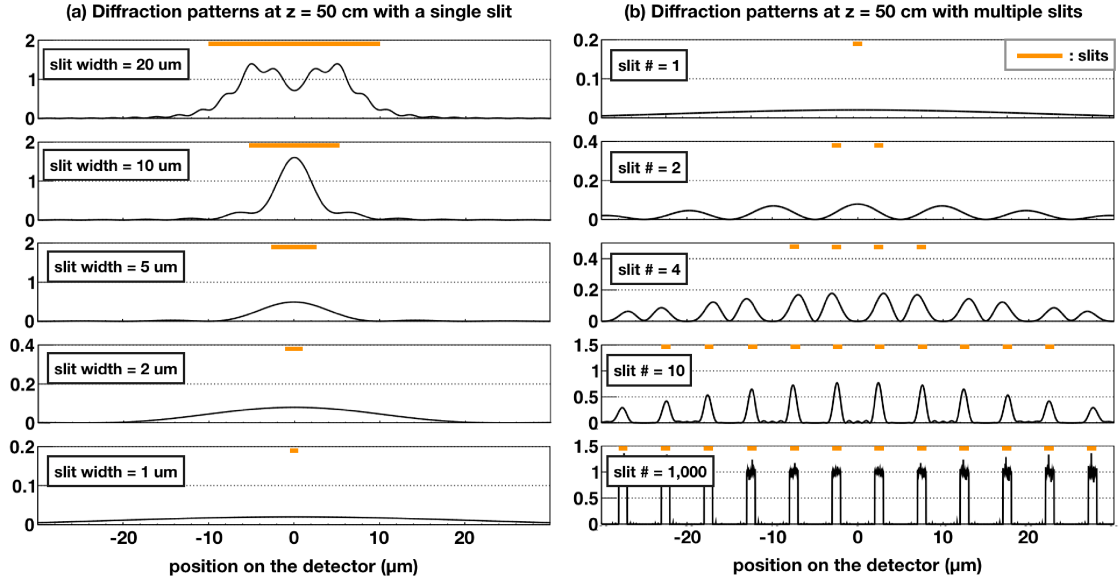}
   \end{tabular}
   \end{center}
   \caption[Diffraction patterns in the cases of a single and multiple slits for monochromatic x-rays.]  
   { \label{fig:simulation_slitN}  
    Calculated diffraction patterns for x-rays with a wavelength of $0.1\nm$, in the cases of (a) a single slit and (b) multiple slits.
    Each panel of (a) and (b) shows a slit width and the number of employed slits, respectively
    (note that incident x-rays have the same intensity for each panel and the ranges of the vertical axes are different between the panels).
    The slit position and width are indicated with orange lines.
    }
   \end{figure}

Interference patterns in the near-field diffraction regime such that Talbot effect occurs can be calculated by means of the Fresnel diffraction equation.
In this section, we present our calculation results of the diffraction patterns for x-rays with a wavelength of $0.1\nm$ in the cases of a single slit and multiple slits.
Figure \ref{fig:simulation_slitN} (a) shows the calculated patterns at $z=50\cm$ performed with a single slit, with the slit width varied from $20\um$ to $1\um$.
Whereas the slit with a width of a few tens of $\si{{\micro}m}$ forms an image even if somewhat affected by a diffraction effect, 
the image with the $1\um$-width slit is highly blurred to $\sim\ang{;;20}$, the result of which indicates that a $1\um$-width slit does not function as a slit camera.
Figure \ref{fig:simulation_slitN} (b) shows the calculated patterns at the same $z$ with equally-spaced multiple slits ($d = 5\um$ and $f = 0.2$), 
with the number of slits varied from 1 to 1000, where the configuration satisfies Eq. \ref{eq:distance_plane} (corresponding to $m$ = 2).
The result clearly demonstrates to what extent the Talbot effect works, depending on the number of slits employed;
self-images are clearly formed as long as the sufficient number of slits are employed, whereas the diffraction is not negligible in cases with only a few slits.

\begin{figure} [ht]
   \begin{center}
   \begin{tabular}{c} 
   \includegraphics[height=8.5cm]{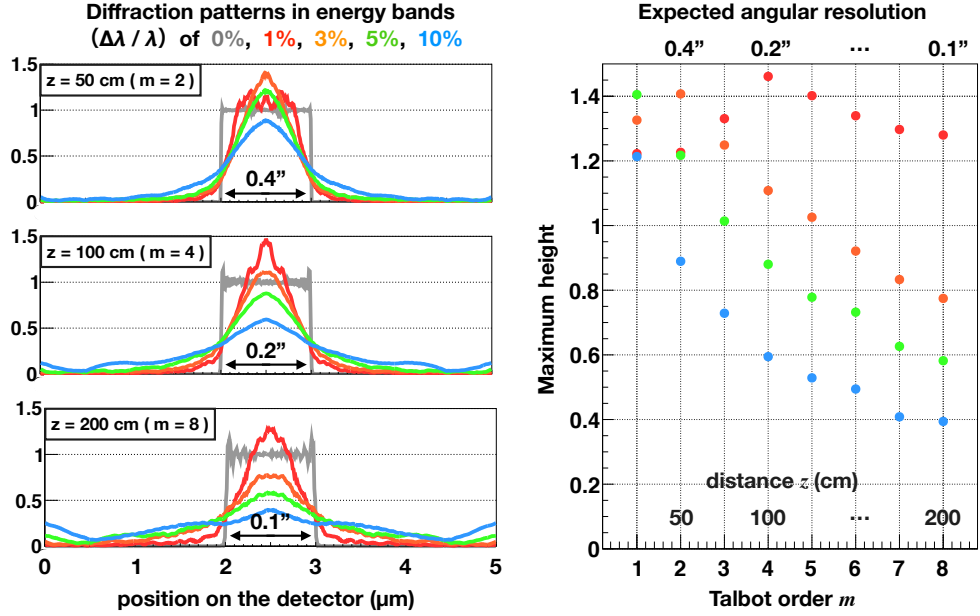}
   \end{tabular}
   \end{center}
   \caption[Diffraction patterns in the case of multiple slits for non-monochromatic x-rays.]  
   { \label{fig:simulation_energy}  
    (Left) Calculated specific diffraction patterns in energy bands of (gray) $0\%$, (red) $1\%$, (orange) $3\%$, (green) $5\%$ and (blue) $10\%$.
    Black arrows indicate the image width calculated according to Eq. \ref{eq:resolution}.
    (Right) Maximum height of the diffraction patterns in the same energy bands as the left panel, as a function of Talbot orders $m$.}
   \end{figure}

In real experiments, we expect that obtained patterns would be somewhat blurred because we conduct imaging with non-monochromatic x-rays 
and the ability to filter out the x-rays out of the target wavelength is limited by the energy resolution of the employed detector.
Then, we calculate the expected image patterns accumulated over energy bands $(\Delta\lambda/\lambda)$ of $0\%$, $1\%$, $3\%$, $5\%$ and $10\%$ centered on $0.1\nm$
for configurations of the number of slits of 1000 with imaging planes at several Talbot orders ($m$=1, 2\,..., 8), varying the wavelength from $0.095\nm$ to $0.105\nm$ in steps of $10^{-4}\nm$.
The left and right panels of Fig. \ref{fig:simulation_energy} represent specific profiles and the maximum height of the image patterns, respectively;
they show that the system with  $z=50\cm$ ($m$ = 2) maintains the angular resolution of $\ang{;;0.5}$ even with the broadest energy band of $10\%$,
though the self-images at any distance increasingly blurs with broader $\Delta\lambda/\lambda$ or higher $m$ unless monochromatic x-rays.
They also imply that employing a detector with a higher energy resolution to MIXIM is important to minimize the blurring.
However, given that the modern standard x-ray detectors of solid-state detectors, including x-ray CCDs and CMOS sensors, 
typically have a reasonably good energy resolution of about 3\%, they should be sufficient for imaging at small $m$.

\section{Proof-of-concept Experiments}
\label{sec:experiment}  

\subsection{Components}
\label{sec:experiment_1}

\begin{figure} [ht]
   \begin{center}
   \begin{tabular}{c} 
   \includegraphics[height=5.5cm]{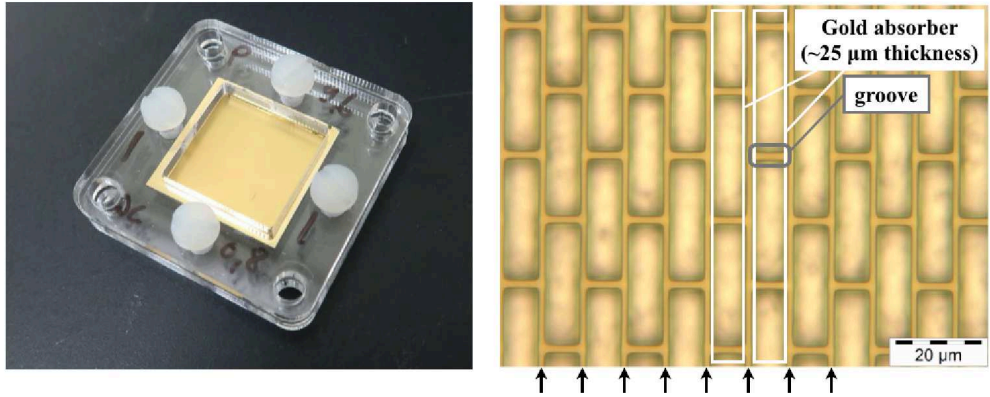}
   \end{tabular}
   \end{center}
   \caption[Figure of the slits used in the experiments.]  
   { \label{fig:Image_grating} 
   (Left) Image of the slits enclosed in an acrylic cover. (Right) Micrograph of the slits. Black arrows show the positions of vertical slits.}
   \end{figure}

The angular resolution of MIXIM is determined by the pitch and opening fraction of multiple slits (section \ref{sec:concepts_1}).
We adopted equally-spaced slits with a pitch of $9.6\um$ and an opening fraction of 0.18 (the same one as that used in our past experiment\cite{Hayashida2018}),
manufactured by the LIGA (the german acronym for lithography, electroplating and molding) process at the Karlsruhe Institute of Technology.
The slit structure is made of gold absorbers with a thickness of about $25\um$ supported by a $0.54\mm$-thick layer of polyimide.
We enclosed the slits in an acrylic cover with a size of $32\mm \times 32\mm$ (left panel of Fig.  \ref{fig:Image_grating}).
The right panel of Fig. \ref{fig:Image_grating} displays a micrograph of the slits, where the vertical slit structure with a pitch of $9.6\um$ is clearly visible 
(we note that the gold absorbers also have similarly-pitched grooves, which are necessary for structural support).

The x-ray detector of MIXIM needs to have a sufficiently high spatial resolution to resolve the expected periodic patterns and sufficiently high energy resolution to extract x-rays with a target wavelength.
In our previous experiment\cite{Hayashida2018}, we adopted fine-pitch slits ($d=9.6\um$, $f =0.18$) and GSENSE5130 (a scientific CMOS sensor developed by Gpixel Inc.).
Since the pixel size of GSENSE5130 ($4.25\um$) was larger than the  slit width ($1.73\um$), resultant spatial resolution was low, hence the low visibility of the self-images.
In this experiment, we introduced the newly available GMAX0505 model, a scientific CMOS sensor released in 2018 with a pixel size of $2.5\um$, 
which is the smallest pixel size among commercially-available scientific CMOS sensors with global shutters.
This sensor has 5120 $\times$ 5120 pixels with an effective imaging area of $12.8\mm \times 12.8\mm$.
Although GMAX0505 was primarily developed as a visible light sensor, we have demonstrated that it can detect x-rays in the photon-counting mode for simultaneous x-ray imaging spectroscopy.
The x-ray spectrum obtained at room temperature has an energy resolution of $176\eV$ at $5.9\keV$\cite{Asakura2019}.
With this spectroscopic capability, we can distinguish detected events originating in an incident x-ray beam, from the background, which is essential for proof-of-concept experiments of MIXIM.

\subsection{Setups}
\label{sec:experiment_2}

\begin{figure} [ht]
   \begin{center}
   \begin{tabular}{c} 
   \includegraphics[height=10.5cm]{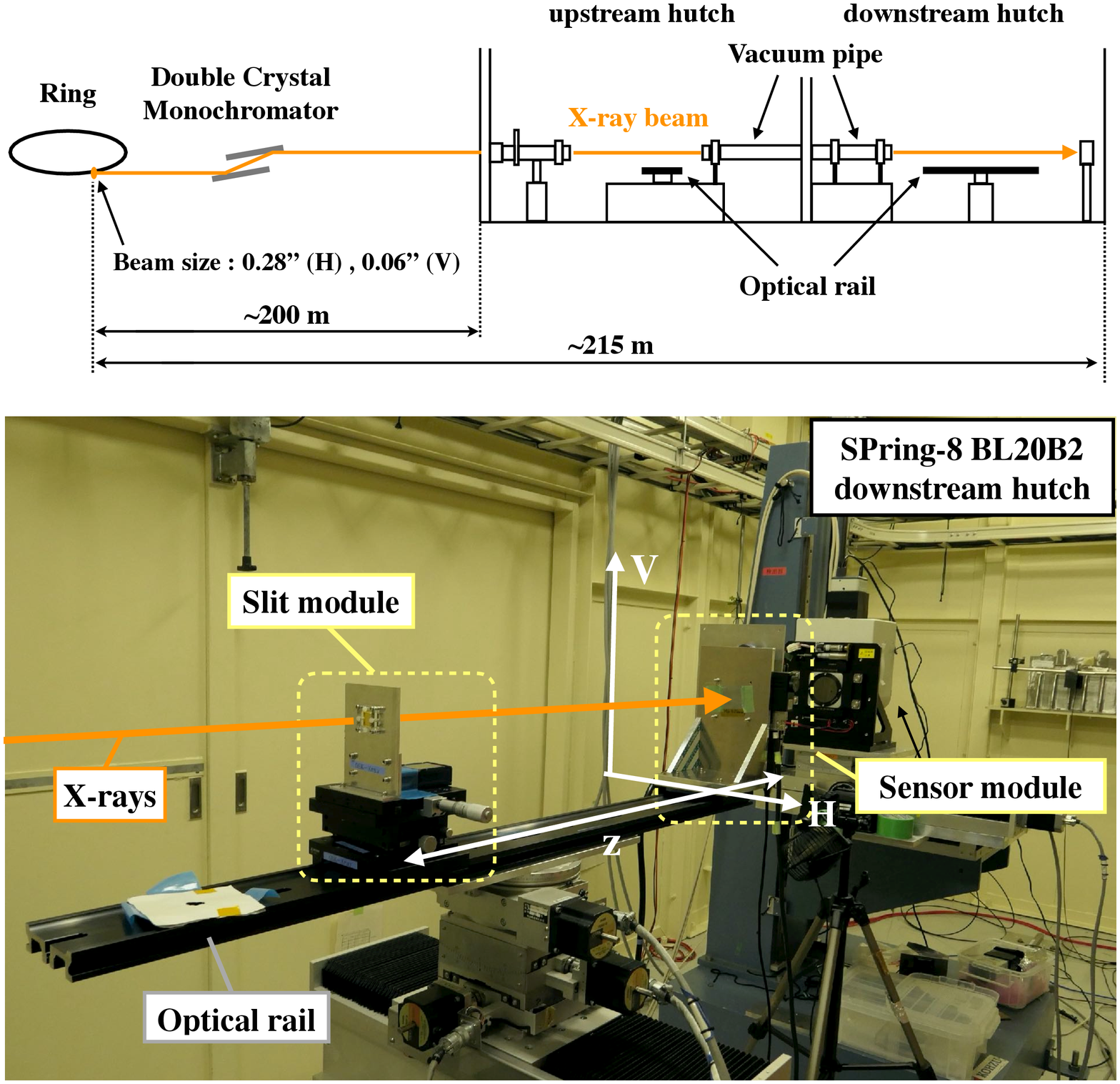}
   \end{tabular}
   \end{center}
   \caption[Overview of SPring-8 BL20B2 and the experiment setup.]  
   { \label{fig:SPring8_setup} 
    (Top) Overview of SPring-8 BL20B2. The slit and sensor modules were installed onto the optical rails at the experiment hutches.
    The beam spot is located about $200\m$ away from the upstream hutch.
    (Bottom) Setup at the downstream hutch during the experiment conducted in 2018 December.}
   \end{figure}

With these components, we conducted proof-of-concept experiments at the beamline BL20B2 of SPring-8, the synchrotron radiation facility in Hyogo, Japan\cite{Goto2001}. 
The top panel of Fig. \ref{fig:SPring8_setup} shows the schematic overview of BL20B2;
the length from the initial beam spot to the end of the downstream hutch is $215\m$, which provides a beam with a high degree of parallelization,
and the x-ray beam was monochromatized with a double crystal monochromator upstream of the experiment hutches,
The beam spot size is $0.29\mm$ (H) and $0.06\mm$ (V), hence the beam divergence of $\ang{;;0.28}$ (H) and $\ang{;;0.06}$ (V);
it should be noted that the vertical spot size is a rough estimate since it includes the fluctuation effect of the crystals in the monochromator.
The beam size was set to be $10\mm \times 10\mm$ to fit within the size of the imaging area of GMAX0505.
Since the x-ray beam at BL20B2 is too high in intensity for our experiments, we inserted an attenuation plate in order to operate the CMOS sensor in the photon-counting mode.

The bottom panel of Fig. \ref{fig:SPring8_setup} shows our actual experimental setup in the downstream hutch.
Optical rails parallel to the beam axis were set in the both hutches.
A slit module and a sensor module were installed on the rail so that the beam would hit both the slits and sensor.
In this setup, the distance between these modules could be easily adjusted by moving them along the rails
(we installed the slit module in the upstream hutch when we extended the distance between these modules beyond the length of the downstream optical rail).
We performed the experiments, including data acquisition, using an evaluation board and a software suite developed and provided by GPixel Inc.,
and processed the obtained frame data with our analysis pipeline explained in section \ref{sec:experiment_3}.

\subsection{Procedures}
\label{sec:experiment_3}

\begin{figure} [ht]
   \begin{center}
   \begin{tabular}{c} 
   \includegraphics[height=8.7cm]{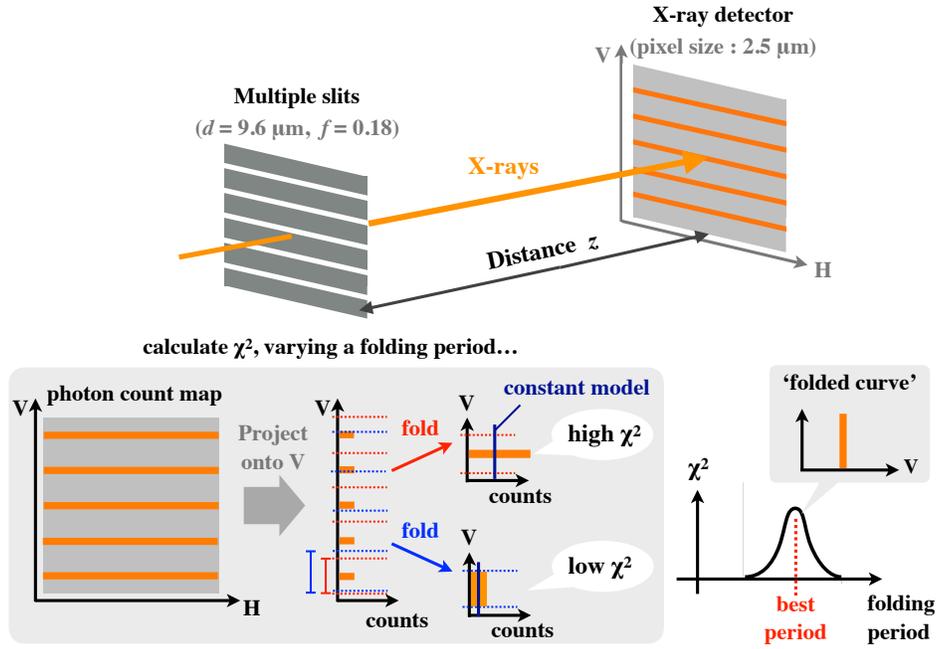}
   \end{tabular}
   \end{center}
   \caption[Schematic chart of analysis procedures of 1D imaging.]  
   { \label{fig:AnalysisFlow_1D} 
    Schematic chart of the analysis procedures of 1D imaging for obtaining the folded curve.}
   \end{figure}

\begin{figure} [ht]
   \begin{center}
   \begin{tabular}{c} 
   \includegraphics[height=8.7cm]{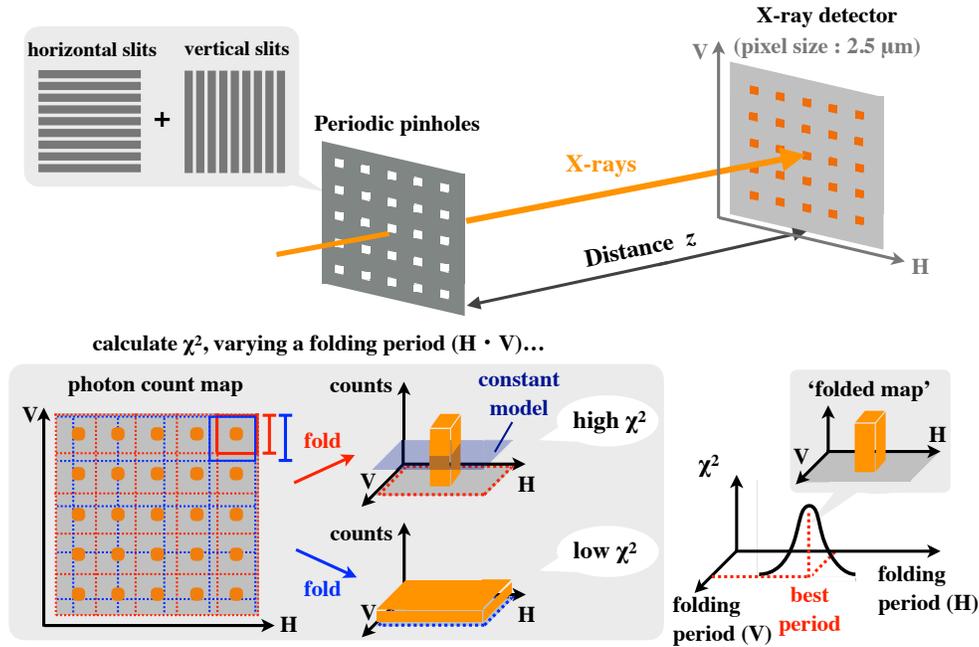}
   \end{tabular}
   \end{center}
   \caption[Schematic chart of analysis procedures of 2D imaging.]  
   { \label{fig:AnalysisFlow_2D}
    Schematic chart of the analysis procedures of 2D imaging for obtaining the folded map.}
   \end{figure}

We first evaluated the performance of one-dimensional (1D) imaging and then conducted two-dimensional (2D) imaging for the first time for MIXIM.
For the 2D imaging, we prepared two uniformly-spaced multiple slits, the same as those mentioned in section \ref{sec:experiment_1}, 
and assembled a periodic pinhole mask by combining them so that their slit directions intersected perpendicularly.
We extracted only x-ray events within the target energy range from the obtained frame data, and generated a photon count map with the events.
Figure \ref{fig:AnalysisFlow_1D} and \ref{fig:AnalysisFlow_2D} show the schematic views of the analysis procedures of the 1D and 2D imaging, respectively.
In the 1D imaging, we generated projected maps along the slit direction and folded it along the axis, varying the folding period.
For each folding period, we fitted the folded profile with a constant model (whose value was fixed to the average counts per bin) and calculated a chi-square, 
assuming that the chi-square value is at the maximum when the folding period best agreed with the periodicity of the self-image.
The analysis procedures with the 2D imaging were almost same as those with the 1D imaging except that we folded the count map two-dimensionally without projection.
Hereafter, we define the folding period with the maximum chi-square as the best-estimate period 
and refer to the 1D and 2D profiles folded with the best-estimate period as the folded curve and folded map, respectively.

\begin{figure} [ht]
   \begin{center}
   \begin{tabular}{c} 
   \includegraphics[height=5cm]{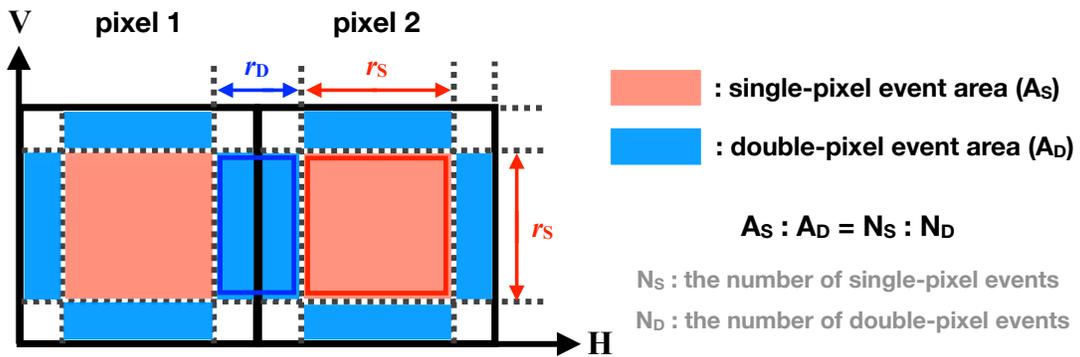}
   \end{tabular}
   \end{center}
   \caption[Definition of pixel areas for sub-pixel analysis technique.]  
   { \label{fig:SubPixel}
   Definition of the areas for single-pixel and double-pixel events in our sub-pixel analysis technique,
   where $r_{\mathrm{S}}$ and $r_{\mathrm{D}}$ are directly calculated from the counting ratio of the single-pixel and double-pixel events.}
   \end{figure}

In order to maximize the spatial resolution of the data obtained with the detector, we also applied a sub-pixel analysis technique before creating the photon count map.
In our analysis, x-ray events extracted from frame data were categorized into three types (see Ref. \citenum{Asakura2019} for detailed description with an image for each event type):
single-pixel events (all x-ray signals reside within a pixel), double-pixel events (signals striding across two pixels), and extended events (signals spread over multiple pixels).
Among these event types, our sub-pixel analysis employs only single-pixel and double-pixel events, discarding the extended events.
Assuming that the spreads of single-pixel events and double-pixel events are located at the center of the pixel and at the boundary of the two pixels, respectively\cite{Tsunemi2001},
we parameterized their incident areas with the respective size parameters $r_{\mathrm{S}}$ and ${r_\mathrm{D}}$ in the way indicated in Fig. \ref{fig:SubPixel} 
with a constraint requiring that their area ratio be equal to the counting ratio between the two event types.
Since the precise incident position of each event could not be identified with this technique, 
we randomized the sub-pixel incident positions of the events of each event type within the corresponding area in our analysis.
In consequence, we derived $r_{\mathrm{S}}$ and $r_{\mathrm{D}}$ to be approximately 0.8 and 0.2 pixels, respectively, at $12.4\keV$ in our experiments.
Thus, our detector has a spatial resolution of $\sim2\um$ for single-pixel events and $\sim0.5\um$ for double-pixel events.

\section{Results}
\label{sec:results}  

\subsection{One-dimensional Imaging}
\label{sec:results_1}

\begin{figure} [ht]
  \begin{center}
  \begin{tabular}{c} 
  \includegraphics[height=5.5cm]{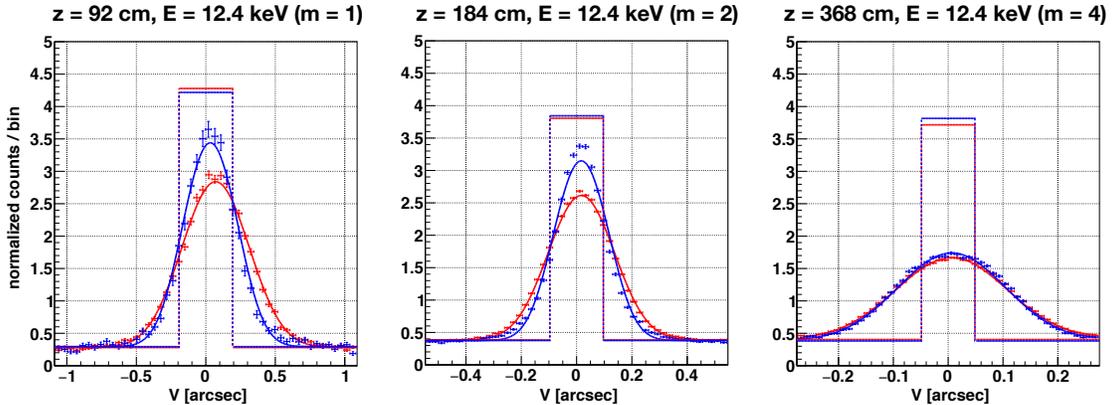}
  \end{tabular}
  \end{center}
  \caption[The results of 1D-imaging.]  
  {\label{fig:Folded_curve} 
    Folded curves at the slit-detector distances (see Fig. \ref{fig:AnalysisFlow_1D}) of $z=92\cm$, $z=184\cm$, and $z=368\cm$ 
    for (red) single-pixel events and (blue) vertically striding double-pixel events.
    The horizontal and vertical axes denote the incident angle and normalized counts per bin.
    The dotted and solid lines are the best-fit square models before and after Gaussian smoothing is applied, respectively.}
\end{figure}

We took frame data for $12.4\keV$ and derived the folded curves at three slit-detector distances $z$: $z=92\cm$ ($m$ = 1), $z=184\cm$ ($m$ = 2), and $z=368\cm$ ($m$ = 4).
Figure \ref{fig:Folded_curve} displays the obtained 50-bin folded curves with single-pixel events and vertically striding double-pixel events at these distances, normalized by the average counts per bin.
The pitch of the slits and the slit-detector distance determine the field of view (FOV) of 1D imaging; 
the FOVs in these cases were derived to be $\ang{;;2.16}$, $\ang{;;1.08}$, and $\ang{;;0.54}$ at $z=92\cm$, $z=184\cm$, and $z=368\cm$, respectively.
Although the folded curves should ideally be square shapes, the actual observed ones were somewhat blurred.
The likely primary causes of the blurring are the beam divergence, fabrication accuracy of the slit structure, and the limited spatial resolution of the detector.

In order to evaluate the imaging performance, we fitted the folded curves with a smoothed square model,
specifically a convolution of a square shape plus a constant and a normalized Gaussian function. 
The square width and normalization of the entire model were fixed at the slit width and total counts, respectively, 
while the square center, the ratio of the constant to the entire model, and the standard deviation of the Gaussian were allowed to vary in the fitting.
We then derived from fitting results the Full Width at Half Maximum (FWHM) and visibility $\mathcal{V}$ to parameterize the actual angular resolution and constant contribution.
In the calculation, we excluded the constant part for the FWHM and defined $\mathcal{V}$ in the following equation:
\begin{linenomath}
\begin{align}
  \label{eq:visibility}
  \mathcal{V} = \frac{C_{\mathrm{max}}-C_{\mathrm{min}}}{C_{\mathrm{max}}+C_{\mathrm{min}}}, 
\end{align}
\end{linenomath}
where $C_{\mathrm{max}}$ and $C_{\mathrm{min}}$ are the maximum and minimum counts per bin of the best-fit model, respectively.
The derived FWHM and $\mathcal{V}$ are tabulated in table \ref{table:1D_Fitting}, 
and the best-fit square models before and after smoothing are overlaid on the observed profiles in Fig. \ref{fig:Folded_curve}.

Notably, the folded curves were more blurred for a larger $m$ though the incident x-rays were almost monochromatized, contrary to the calculation in section \ref{sec:concepts_2}.
The main reason is that the parallel beam approximation used for deriving $z$ are increasingly less appropriate for larger $z$.
The simple equation Eq. \ref{eq:distance_plane}, which assumes the ideal case of zero divergence, is corrected for the spherical-wave beam with a non-zero divergence in the following formula:
\begin{linenomath}
\begin{align}
  \label{eq:distance_sphere}
  z_{\mathrm{T}} = m\frac{d^2}{\lambda} \frac{z_0}{z_0-\frac{md^2}{\lambda}} \; (m = 1, 2, 3...),
\end{align}
\end{linenomath}
where $z_0$ represents the distance between the x-ray source and slits\cite{Patorski1988}.
It follows that the optimal distance $z$ for a spherical wave is larger than that derived with Eq. \ref{eq:distance_plane}.
The degree of the effect was observable in this experiment, resulting in the deterioration of the quality of the self-imaging.
Nevertheless, even the folded curve at the largest distance $z=368\cm$ still retained a visibility of more than 0.5, which was reasonably good.
Consequently, these results show that MIXIM, of which the system size is less than $1\m$, has a similar angular resolution to that of \textit{Chandra}
and technically can realize an even higher angular resolution than that by increasing the distance $z$.

\begingroup
\renewcommand{\arraystretch}{1.3}
\begin{table}[ht]
\caption[Summary of the 1D fitting results.]{Summary of the fitting results. Ideal values are simply derived with Eq. \ref{eq:resolution}.}
\label{table:1D_Fitting}
\begin{center}
\begin{tabular}{ccccccc}
        \toprule
        \multicolumn{5}{c}{Configuration}                                                                  &  \multicolumn{2}{c}{Performance index}                                  \\ \hline
        $z$ (\cm)  &  $E$ (\keV) &  $m$   &     event type    &         ideal value        &             $\mathcal{V}$              &        FWHM (arcsec)             \\ \midrule
             92        &      12.4       &     1     &   single-pixel     &       $\ang{;;0.39}$     &   $0.821^{+0.003}_{-0.003}$   &  $0.571^{+0.006}_{-0.006}$  \\  
                         &                    &            &   double-pixel   &       $\ang{;;0.39}$     &    $0.843^{+0.006}_{-0.006}$   &  $0.461^{+0.007}_{-0.007}$  \\  
             184      &      12.4       &     2     &   single-pixel     &      $\ang{;;0.19}$     &    $0.745^{+0.001}_{-0.001}$   &   $0.282^{+0.002}_{-0.002}$ \\
                         &                    &            &   double-pixel   &       $\ang{;;0.19}$     &    $0.788^{+0.002}_{-0.002}$   &   $0.232^{+0.002}_{-0.002}$ \\
             368      &      12.4       &     4     &   single-pixel    &      $\ang{;;0.097}$     &   $0.568^{+0.002}_{-0.002}$   &  $0.254^{+0.003}_{-0.003}$   \\
                         &                    &            &   double-pixel   &      $\ang{;;0.097}$    &    $0.606^{+0.003}_{-0.003}$   &  $0.241^{+0.003}_{-0.003}$   \\                           
      \bottomrule
    \end{tabular}
    \end{center}
\end{table}
\endgroup

\begin{figure} [ht]
   \begin{center}
   \begin{tabular}{c} 
   \includegraphics[height=6.5cm]{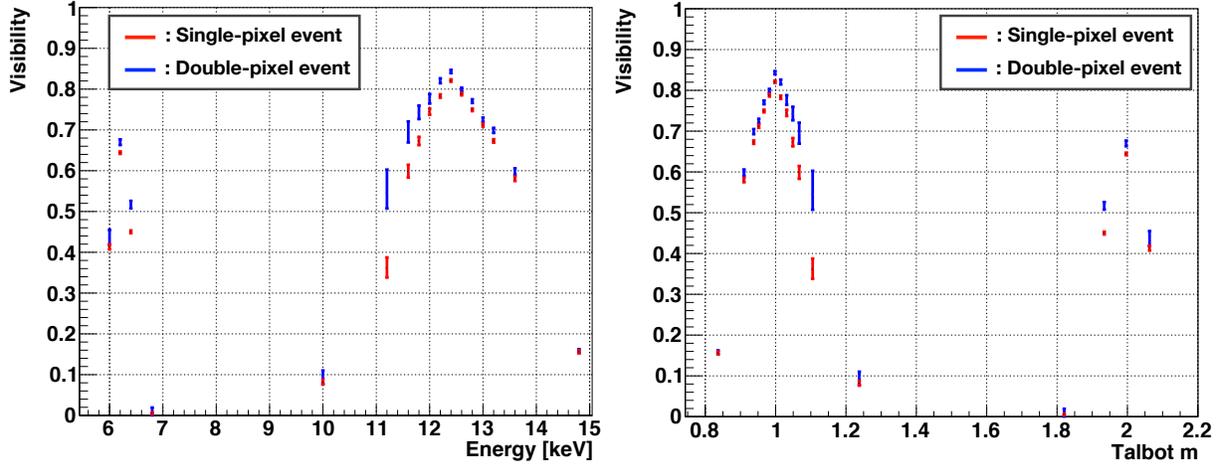}
   \end{tabular}
   \end{center}
   \caption[Energy dependence of the visibility.]  
   { \label{fig:E-Visivility_1D} 
   (Left) Energy dependence of the visibility with $z=92\cm$ and $d=9.6\um$. (Right) Same as the left panel, but plotted along the horizontal axis of $m$.}
   \end{figure}

We also irradiated x-rays in a variety of energies, ranging from $6\keV$ to $15\keV$, to the detector placed at $z = 92\cm$ to investigate the energy dependence of the visibility.
Figure \ref{fig:E-Visivility_1D} shows the obtained visibilities from the results.
The visibilities were peaked at roughly $12.4\keV$ ($m$ = 1) and $6.2\keV$ ($m$ = 2), and were higher than 0.5 within an energy band of $\pm(10/m)\%$ around these peaks.
The energy band extension is in fact a trade-off with the imaging performance, yet a wide energy band serves as a great advantage in practical space observations, 
considering that long exposure times are required for imaging with high photon statistics since photon fluxes of cosmic x-ray sources are usually low.

\subsection{Two-dimensional Imaging}
\label{sec:results_2}

\begin{figure} [ht]
   \begin{center}
   \begin{tabular}{c} 
   \includegraphics[height=6.5cm]{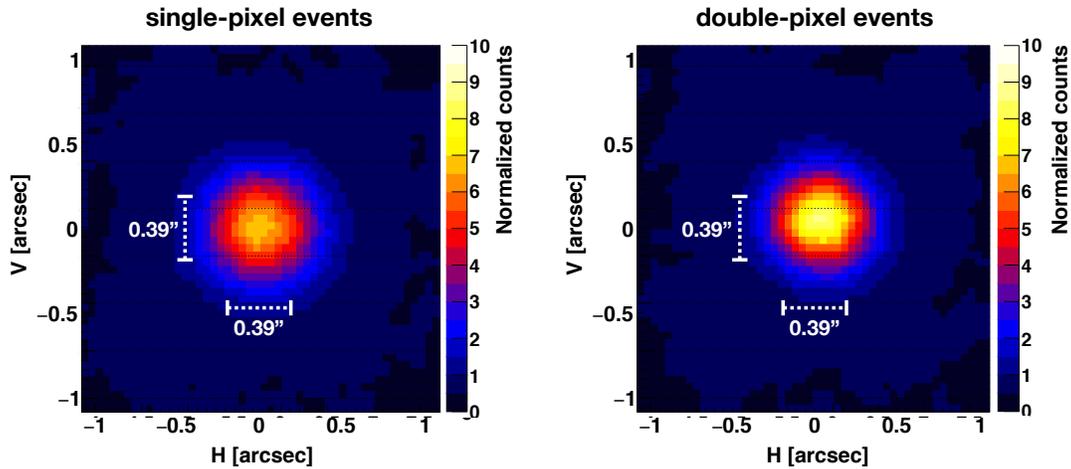}
   \end{tabular}
   \end{center}
   \caption[The results of 2D-imaging at $z=92\cm$.]  
   { \label{fig:Folded_map_92cm} 
    Folded maps at $z=92\cm$ with single-pixel events (left) and double-pixel events (right), smoothed with a Gaussian for better visualization.
    The color scale shows the normalized counts and the dotted white lines denote the theoretical value in the ideal case according to Eq. \ref{eq:resolution}.} 
   \end{figure}

To evaluate the quality of 2D imaging with our system, we first conducted an experiment, setting $(E, z)$ to $(12.4\keV, 92\cm)$.
Figure \ref{fig:Folded_map_92cm} shows the obtained folded maps (binned with $50\times50$ pixels) with single-pixel and double-pixel events.
Both the folded maps clearly show that we succeeded in obtaining a 2D profile of the x-ray beam.
We notice that the folded map with double-pixel events is significantly elongated along the horizontal axis, which gives a hint of the intrinsic beam divergence.
As with the case of 1D imaging, the FOV in this case was calculated to be $\ang{;;2.16} \times \ang{;;2.16}$ at $z=92\cm$.

\begin{figure} [ht]
   \begin{center}
   \begin{tabular}{c} 
   \includegraphics[height=6cm]{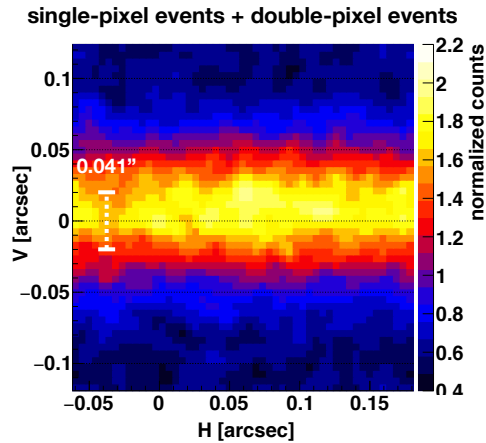}
   \end{tabular}
   \end{center}
   \caption[The results of 2D-imaging at $z=866.5\cm$.]  
   { \label{fig:Folded_map_867cm} 
    Same as Fig. \ref{fig:Folded_map_92cm}, but this shows the combined map with single-pixel and double-pixel events both obtained at $z=866.5\cm$.}
   \end{figure}

Then we extended the distance $z$ from $92\cm$ ($m$ = 1) to $866.5\cm$ ($m$ = 9), the maximum distance achievable in the experiment hutches, 
with which the angular resolution was expected to improve considerably, and conducted another experiment
(in this experiment, we used Eq. \ref{eq:distance_sphere} to derive the relation between $z$ and $m$, taking into account the effect of the beam divergence).
Figure \ref{fig:Folded_map_867cm} shows the obtained folded maps with the same binning as the above-mentioned ones with both single-pixel and double-pixel events.
The result demonstrates that the intrinsic beam profile can be successfully resolved thanks to the very high angular resolution, 
whereas the improvement of the angular resolution is in return for the narrow FOV since the elongation of the distance $z$ corresponds to the zooming-in.
Notably, the horizontal beam divergence ($\ang{;;0.28}$) is more extended than the FOV ($\ang{;;0.24}$) at $z=866.5\cm$,
where the outermost part of the image protrudes from one side of the FOV and contaminates the other side, because of the folding analysis procedure (hereafter we refer to it as the wrap-around effect).

To evaluate the imaging performance, we applied the two-dimensional version of the 1D smoothed square model (section \ref{sec:results_1}): 
a convolution of a square box plus a constant plane with a normalized 2D Gaussian function.
The horizontal and vertical standard deviations of the 2D Gaussian and the ratio of the constant plane were allowed to vary in the fitting,
while the other parameters were fixed in the same way as in the 1D-model fitting. The wrap-around effect was taken into account in the model.
We obtained the horizontal and vertical FWHM to be $\ang{;;0.52}$ and $\ang{;;0.47}$ at $z=92\cm$, which are roughly agreed with the result of the 1D imaging.
As for the results for $z=866.5\cm$, the vertical FWHM was measured to be less than $\ang{;;0.1}$,
whereas the horizontal FWHM was failed to be determined due to its almost flat profile.
Table \ref{table:2D_Fitting} summarizes the fitting results.

Finally, one more important point to consider is the effect of the intrinsic beam divergence of $\ang{;;0.06}$.
Since the value is comparable with the derived angular resolution, our above-derived value should be a considerable overestimate for the true angular resolution of the system.
Taking the effect of the intrinsic beam divergence into account, we obtained a better estimate for it to be $\ang{;;0.05}$.

\begingroup
\renewcommand{\arraystretch}{1.3}
\begin{table}[ht]
\caption[Summary of the 2D fitting results.]{Summary of the fitting results of our 2D imaging experiments.}
\label{table:2D_Fitting}
\begin{center}
\begin{tabular}{ccccccc}
        \toprule
        \multicolumn{4}{c}{Configuration}                                &  \multicolumn{3}{c}{Performance index}                                                                                                    \\ \hline
        $z$ (\si{cm})  &  $E$ (\si{keV}) &  $m$   &     event type       &              $\mathcal{V}$            &           H-FWHM (arcsec)             &        V-FWHM (arcsec)           \\ \midrule
                92          &         12.4         &     1     &    single-pixel       &   $0.858^{+0.002}_{-0.002}$  &      $0.565^{+0.002}_{-0.002}$    &   $0.534^{+0.002}_{-0.001}$  \\
                              &                         &            &    double-pixel      &   $0.884^{+0.002}_{-0.002}$  &     $0.515^{+0.001}_{-0.003}$     &   $0.469^{+0.001}_{-0.002}$  \\
             866.5        &         12.4         &     9     &   single+double    &   $0.522^{+0.011}_{-0.011}$  &                     N/A                          &    $0.080^{+0.001}_{-0.001}$  \\
      \bottomrule
    \end{tabular}
    \end{center}
\end{table}
\endgroup

\section{Discussion}
\label{sec:discussion}  

\subsection{Comparison with Past Results}
\label{sec:discussion_1}

The folded curves and folded maps at $z=92\cm$ ($m$ = 1) have a FWHM of about $\ang{;;0.5}$ and a visibility of more than 0.8,
which are certainly improved from those of our previous experiment\cite{Hayashida2018}.
Since the previous experiment employed the same slits as this one, the performance improvement was certainly attributed to the difference in the detector, more specifically the improved spatial resolution.
In addition, we found that the results with double-pixel events were more useful in terms of better FWHMs and visibilities than those with single-pixel events, yielding higher spatial resolutions for the events at $12.4\keV$.
Notably, the double-pixel events at $m$ = 1 had widths wider than theoretical values in the ideal case according to Eq. \ref{eq:resolution}
even after subtracting the intrinsic beam divergence, which may suggest that the further improvement of the spatial resolution enhances the angular resolution.
We also notice that the visibility did not reach unity due to the constant component, the same as with the previous experiment.
Although the cause of the constant component is not yet thoroughly understood, it might be due to x-rays passing through the grooves described in section \ref{sec:experiment_1}.

\subsection{Notable Features of MIXIM}
\label{sec:discussion_2}

In interferometers in general, the maximum achievable angular resolution is an increasing function of the system size.
However, a large size typical to large interferometry projects means less feasibility in practice.
It is particularly the case in the field of x-ray interferometry, which may require space formation flight of multiple satellites 
maintaining an ultra-high precision to realize a hundreds of kilometers long baseline of the system (section \ref{sec:intro}).
MIXIM is a scalable imaging system and can retain a reasonable capacity even when the system size is small, as demonstrated in the present experiment;
a small prototype MIXIM with a size smaller than $1\m$ showed an angular resolution of $\sim\ang{;;0.5}$ which is comparable to that of \textit{Chandra} and is very good.
Or, MIXIM can be configured to an even smaller size, which would be handy for pioneer in-orbit flight tests, if with some sacrifice in the angular resolution.
Hence, the feasibility for the real deployment of MIXIM in orbit is high.
We should note that employing a detector with a high spatial resolution can be beneficial also for other proposed x-ray interferometers\cite{Noda2022};
the size of the $100\m$ long \textit{MAXIM} prototype can be reduced by employing one, though the developer team has to resolve the inherent problem of fringe spacing.

MIXIM, being a scalable imaging system, can be scaled up to a size of a large satellite or larger, if opportunities arise.
In our estimate, MIXIM with a distance of $z=100\m$ is expected to have an angular resolution of $\ang{;;0.01}$ for $12.4\keV$ with $d\sim70\um$ and $f=0.1$ ($m$ = 2).
If higher $m$ is adopted, the angular resolution for the same slits and wavelength will be even more enhanced though a longer distance will be required 
and a high energy resolution of the detector will be crucial to maintain the imaging quality, as our estimate in section \ref{sec:concepts_2} shows.
A decrease of an opening fraction also leads to improve the angular resolution, 
whereas high spatial resolution is simultaneously required since it determines the maximum element number in a given image of MIXIM.
In particular, a real advantage of MIXIM is potential use in a higher-energy x-ray band; 
whereas such high-energy x-rays are difficult to focus with conventional x-ray mirrors, MIXIM is well capable of achieving high-resolution imaging in such a high energy band
by employing fine-pitch slits with thick absorbers and a detector with a high spatial resolution and high detection efficiency.

While MIXIM has an excellent angular resolution, it also has some constraints originating from the imaging principle.
One of the main constraints is the limited size of FOV (approximately $dz^{-1}$), which will be a significant point to consider once MIXIM has been deployed for space observations.
The inherently narrow FOV of MIXIM due to a small $d$ is one thing but the associated wrap-around effect is another 
because the effect increases the background level in particular when other x-ray sources reside outside the FOV or when the target source has a more extended structure than the FOV.
Thus, MIXIM's observational targets should be almost point-like at a wavelength of interest.
Notably, an increase of the pitch $d$ practically leads to an even narrower FOV because it also requires the extension of the distance $z$ 
to maintain the X-ray interferometry with the same $m$ value according to Eq. \ref{eq:distance_plane} ($z \propto d^2$).
Hence, to achieve simultaneously a high angular resolution and wide FOV, the system size and opening fraction are both required to be reduced, which results in an accordingly reduced effective area.

\subsection{Application to X-ray Astronomy}
\label{sec:discussion_3}

Finally, we briefly consider potential applications of MIXIM to x-ray astronomical observations.
In space observations, MIXIM requires an additional collimator to block x-rays coming from the outside of the FOV,
because otherwise the wrap-around effect of other x-ray sources (including diffuse sources, such as the cosmic x-ray background) would inevitably contaminate the image.
Although the perfect elimination of contamination is unfeasible for MIXIM, a parallel hole collimator can be used to reduce the background;
if we install fine-pitch slits onto a collimator with a length of $120\cm$ and a hole size of $1\cm$, x-rays with incident angles of more than $\ang{1}$ are blocked.
We should note, however, that it is intrinsically difficult for MIXIM to capture the accurate structure of diffuse x-ray sources extended more than the FOV even if MIXIM is equipped with an effective collimator.

We also point out that MIXIM has a low sensitivity because it does not have a focusing capability.
Inferred from past collimator instruments, such as \textit{Ginga} LAC\cite{Hayashida1989}, 
the observational targets of MIXIM should be brighter than roughly 1 mCrab ($2.4\times10^{-11}\mathrm{\,erg\,s^{-1}\,cm^{-2}}$ in 2--10$\keV$) as a conservative estimate.
In reality, the limiting sensitivity might be improved by employing an effective method to reduce the non-x-ray background, such as event grade selection or the installation of active shields.
The effective area of MIXIM should be also increased from the current prototype system which has a low opening fraction,
although unbound increase of the effective area to hundreds or thousands $\si{cm^2}$ is almost pointless 
because only bright x-ray sources are targets of MIXIM anyway because of its background-limited sensitivity.

Taking these constraints into account, we propose nearby active galactic nuclei (AGN) as one class of the most interesting candidate targets.
Whereas most AGNs have been conventionally regarded as point-sources for the current x-ray astronomical satellites,
high-resolution x-ray imaging with \textit{Chandra} resolved the profiles of some nearby AGNs extending a few arcseconds in the iron K$\alpha$ band
(e.g., Circinus galaxy\cite{Smith2001}, NGC1068\cite{Young2001}, and NGC4945\cite{Marinucci2012}). 
It is reasonable to guess that there are many other AGNs like them, which have been regarded as point-sources 
simply because of the limitation of the past x-ray observatories but in fact have unresolved fine spatial structures.
Such potentially spatially-extended AGNs are one of the most suitable observational targets for MIXIM.
High-resolution imaging with MIXIM will provide a new perspective for these classes of astrophysical x-ray sources.

\begin{figure} [ht]
   \begin{center}
   \begin{tabular}{c} 
   \includegraphics[height=6cm]{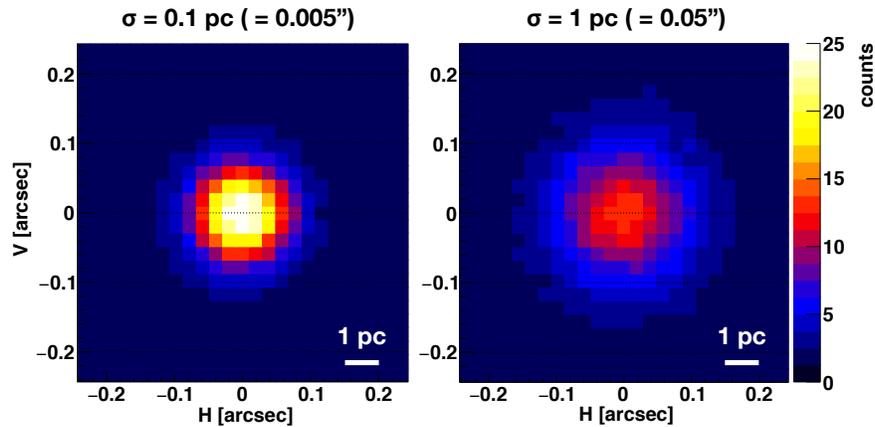}
   \end{tabular}
   \end{center}
   \caption[Expected images of the Circinus galaxy observed with MIXIM for 10\,Ms.] 
   { \label{fig:Circinus}
   Expected images of the Circinus galaxy observed in the iron K$\alpha$ band with MIXIM ($z=4.3\m$) for 10\,Ms,
   assuming that the iron K$\alpha$ emission has a Gaussian profile with a standard deviation of (left) 0.1\,pc and (right) 1\,pc, respectively.}
   \end{figure}

Finally, we concisely discuss specific requirements in the case of an observation of the Circinus galaxy as an example of candidate targets.
\textit{XMM-Newton} obtained the iron K$\alpha$ photon flux from the Circinus galaxy to be $2.14 \pm 0.06 \times 10^{-4}\mathrm{\,photons\,s^{-1}\,cm^{-2}}$\cite{Molendi2003},
which is an order of magnitude higher than the CXB flux in the same energy band (if with the aforementioned parallel hole collimator).
Given that an opening fraction of a mask and effective exposure time is 3.2\% and 10\,Ms ($\sim$4 months), respectively, 
MIXIM necessitates a detector with an effective area of $\sim30\mathrm{\,cm^{2}}$ to collect $2\times10^3$ photons;
while hundreds of sensors are required if with GMAX0505 due to its thin detection layer,
it would be deployable on a single satellite if we adopt a sensor with both a high spatial resolution and high detection efficiency in the future.
Hence, we consider a relatively compact imaging system with $f=0.18\times0.18$, $d=9.6\um$ and $z=4.3\m$ (i.e., $m=9$ for $6.4\keV$ x-rays);
such a system would have an angular resolution of $\sim\ang{;;0.1}$,
which enables us to investigate the spatial structure of the Circinus galaxy (with a distance of 4.2\,Mpc\cite{Freeman1977}) with a pc-scale resolution.
Since the point spread function (PSF) of this configuration could be estimated from the vertical profile of the obtained self-image at $z=866.5\cm$
(from which the beam divergence of $\ang{;;0.06}$ was deducted) with the assumption that the angular resolution is proportional to the mask-sensor distance,
we simulated the observation with an exposure time of 10\,Ms by simply distributing $2\times10^3$ photons according to the PSF convolved with a given source profile.
The left and right panels of Fig. \ref{fig:Circinus} show expected images when an x-ray source has a Gaussian profile with a standard deviation of 0.1\,pc and 1\,pc, respectively.
Whereas we should take the effect of background components and systematic errors into account for further detailed discussion, 
they suggest that MIXIM has a potential to distinguish whether the iron K$\alpha$ emission distributes over a few pc or within the inner sub-pc.

\section{Summary}
\label{sec:summary}  

For realization of unprecedentedly high-resolution x-ray imaging, we have developed MIXIM: a novel x-ray imaging system composed of fine-pitch uniformly-spaced slits and an x-ray detector.
In the case of a simple slit camera, improvement of the angular resolution by reducing the slit width has a serious limit, i.e., the diffraction limit.
MIXIM overcomes the limitation with utilization of the Talbot effect and realizes a high angular resolution for a single wavelength.
In our past proof-of-concept experiments of one-dimensional imaging, spatial resolution of the employed x-ray detector was limited to $4.25\um$, resulting in a still limited angular resolution at the time.
In the present experiment, we introduced a CMOS sensor with a pixel size of $2.5\um$ and conducted proof-of-concept experiments of MIXIM at SPring-8 BL20B2.
The experiment of 1D imaging demonstrated that MIXIM has an angular resolution of about $\ang{;;0.5}$ at $12.4\keV$ with a system size of less than $1\m$
and maintains high visibility even if the chosen energy band is as broad as $\pm$(10/$m$)\% around a target energy, where the visibility is an inversely-increasing function of the energy-band width.
We also conducted two-dimensional imaging, replacing the multiple slits used for 1D-imaging with a multiple-pinhole mask with uniform spacing and successfully obtained 2D x-ray images for the first time with MIXIM.
With 2D-experiments with varying distances between the mask and detector $92\cm$ to $866.5\cm$, we demonstrated that MIXIM obtained an angular resolution of less than $\ang{;;0.1}$.
Notably, the size of MIXIM is adjustable from a micro-satellite size to formation-flying according to the required performance.
An in-orbit small-scale prototype mission is probably more feasible with MIXIM than with the other proposed x-ray interferometers, although MIXIM still has some technical issues.
As interferometers have explored new fields in a variety of wavelengths, MIXIM is capable of resolving yet veiled, very fine spatial structures of x-ray sources in the future with an unprecedentedly-high angular resolution.

\subsection* {Acknowledgments}

The authors are grateful to T. Hanasaka, K. Okazaki, S. Sakuma, A. Ishikura, M. Hanaoka, S. Ide, K. Hattori (Osaka University), 
Dr. M. Hoshino and Dr. K. Uesugi (JASRI) for their support with the proof-of-concept experiments.
This work was supported by Japan Society for the Promotion of Science (JSPS) KAKENHI Grant Numbers 
20J20685 (KA), 18K18767, 19H00696, 19H01908, 20H00176 (KH), 19K21884 (HN), 18H01256, 20KK0071 (HN).
KH was supported by the Mitsubishi Foundation Research Grants in the Natural Sciences 201910033. 
The synchrotron radiation experiments were performed at the BL20B2 of SPring-8 with the approval of the Japan Synchrotron Radiation Research Institute (JASRI) 
with Proposal No. 2018A1368, 2018B1235, 2019A1503, 2019B1492, 2020A1506 and 2021A1442 (KH).
A part of the results of this manuscript was previously reported in SPIE proceedings\cite{Asakura2020}.


\bibliography{report}   
\bibliographystyle{spiejour}   


\vspace{2ex}\noindent\textbf{Kazunori Asakura} is a graduate student at Osaka University in Japan. 
He received his BS and MS degrees from Osaka University in 2018 and 2020, respectively.
His current research interests include X-ray astronomy and instrumentation for X-ray imaging spectroscopy and polarimetry.

\vspace{1ex}
\noindent Biographies and photographs of the other authors are not available.

\listoffigures
\listoftables

\end{spacing}
\end{document}